\documentclass[preprint,showpacs,preprintnumbers,amsmath,amssymb]{revtex4}
\usepackage{epsfig}
\usepackage{graphicx}% Include figure files
\usepackage{dcolumn}% Align table columns on decimal point
\usepackage{bm}% bold math
\usepackage{amsmath}
\usepackage{subfigure}
\usepackage{color}

\let\jnfont=\rm
\def\NPB#1,{{\jnfont Nucl.\ Phys.\ B }{\bf #1},}
\def\PLB#1,{{\jnfont Phys.\ Lett.\ B }{\bf #1},}
\def\EPJC#1,{{\jnfont Eur.\ Phys.\ Jour.\ C }{\bf #1},}
\def\PRD#1,{{\jnfont Phys.\ Rev.\ D }{\bf #1},}
\def\PRL#1,{{\jnfont Phys.\ Rev.\ Lett.\ }{\bf #1},}
\def\MPLA#1,{{\jnfont Mod.\ Phys.\ Lett.\ A }{\bf #1},}
\def\JPG#1,{{\jnfont J.\ Phys.\ G}{\bf #1},}
\def\CTP#1,{{\jnfont Commun.\ Theor.\ Phys.\ }{\bf #1},}
\def\ZPC#1,{{\jnfont Z.\ Phys.\ C }{\bf #1},}
\def\JHEP#1,{{\jnfont JHEP \ }{\bf #1},}
\def\Rv{\not{\hbox{\kern-1pt $R$}}}
\def\p{\not{\hbox{\kern-3pt $p$}}}

\newcommand{\bc}{\begin{center}}
\newcommand{\ec}{\end{center}}
\newcommand{\gev}{{~\text{GeV}}}

\begin{document}
\preprint{\parbox{1.2in}{\noindent arXiv:}}

\title{QCD corrections to the production of $t\bar{t}\gamma$ at the ILC}

\author{Ning Liu\\~ \vspace*{-0.3cm}}
\affiliation{Physics Department, Henan Normal University, Xinxiang
453007, China \vspace*{1.5cm}}

\begin{abstract}
A precise calculation of the top quark pair production associated
with a hard photon is essential for testing the electroweak property
of the top quark in the Standard Model (SM). We investigate the
one-loop QCD corrections to the process $e^{+}e^{-} \to
t\bar{t}\gamma$ at the International Linear Collider (ILC), and find
that the $K$-factor can be as large as 1.238~(1.105,~1.060) for a
center-of-mass energy $\sqrt{s}=500~(800,~1500)$ GeV. The transverse
momentum distributions of the top quark and photon are respectively
shown at leading order (LO) and next-to-leading order (NLO). Due to
the asymmetric rapidity distribution of the top (anti-top) quark, we
also study the top quark forward-backward asymmetry ($A^{t}_{FB}$)
in $t\bar{t}\gamma$ production at NLO, which is found to be
$45.82$~($55.25$,~$55.89$)\% for $\sqrt{s}=500~(800,~1500)$ GeV.
\end{abstract}


\maketitle

\section{INTRODUCTION}
The Standard Model has been well testified by various experiments
\cite{SM}, except that the Higgs scalar is still left as a missing
piece. Therefore, testing the mechanism of electroweak symmetry
breaking (EWSB) \cite{higgs} is now an urgent task for particle
physics. The top quark, which was discovered at the Tevatron in 1995
\cite{top-exp1}, is distinguished for its large mass and short
lifetime \cite{top-exp2}. Since it is free from the QCD confinement,
the top quark productions and decays are much cleaner than the light
quarks \cite{top-decay}. Thus, the top quark is speculated to be a
sensitive probe for the EWSB and new physics \cite{top-rev}. So far,
except for the forward-backward asymmetry in $t\bar{t}$ production
at the Tevatron \cite{top-afb}, most of the measurements on the top
quark are consistent with the Standard Model (SM) predictions, such
as the cross sections of the $t\bar{t}$ production and the single
top production \cite{top-exp3}.

Very recently, the CDF Collaboration has reported its observation of
$t\bar{t}\gamma$ events with a luminosity of 6.0 fb$^{-1}$
\cite{CDF-ttr}. However, the small statistics still limits the
precision study of the gauge coupling of $t\bar{t}\gamma$. In
addition, since the initial photon radiation severely affects the
sensitivity of $t\bar{t}\gamma$ production to anomalous top quark
couplings \cite{top-exotic-charge}, the SM electric charge of the
top quark has not been measured directly up to now. Although the
dominant contribution to the $t\bar{t}\gamma$ production comes from
the gluon fusion at the LHC, it is challenging to determine the top
charge by measuring the cross section ratio
$\sigma(t\bar{t}\gamma)/\sigma(t\bar{t})$ \cite{LHC-ttr} due to the
huge QCD backgrounds. The LO and NLO QCD calculations of the process
$p p(\bar{p}) \to t\bar{t}\gamma$ have been recently carried out at
hadron colliders in Ref.\cite{ttr-hadron}.

In contrast, as a clean top quark factory, the ILC will allow for a
precison test for the top quark property \cite{top-ilc}. Since the
gauge coupling of $t\bar{t}\gamma$ is sensitive to new physics, it
has been studied intensively in a model independent way
\cite{top-gauge-indep}. Some new physics models can also affect the
$t\bar{t}\gamma$ coupling sizably, such as the Little Higgs model
\cite{top-lh}. It is also found that in the supersymmetric and
multi-Higgs models \cite{top-cp} a sizable top quark electric(weak)
dipole momentum can be induced by the non-standard CP violating
interactions. At the ILC, due to a high luminosity, such anomalous
couplings of the top quark can be measured at the one percent level,
which is much better than that at the LHC \cite{top-ilc}. Therefore,
the high order calculations for the top quark processes at the ILC
are needed to meet the experimental precision. The QCD and
electroweak corrections to the process $e^{+}e^{-},\gamma\gamma \to
t \bar{t}$ have been studied in Ref.\cite{tt-ilc}. The CP violation
effects induced at loop level in the top pair production through
(un)polarized $e^{+}e^{-}$ annihilation \cite{ttcp-ee} or photon
fusion \cite{ttcp-rr} have been investigated in some extensions of
the SM. In Ref.\cite{bbr-lep}, the authors have studied the QCD
corrections to the $b\bar{b}\gamma$ production at LEP1. In this
paper, we calculate the one-loop QCD corrections to the process
$e^{+}e^{-} \to t \bar{t} \gamma$ at the ILC.

This work is organized as follows.  In section II, a brief
description for the NLO QCD calculations is given. The
discussions and numerical studies are presented in section III.
Finally, the conclusions are drawn in section V.

\section{A description of analytical calculations}

In our calculations, the NLO QCD corrections ($\Delta\sigma_{QCD}$)
are divided into two parts: the virtual corrections
($\Delta\sigma_{vir}$) and the real gluon radiation corrections
($\Delta\sigma_{real}$). We adopt the dimensional regularization to
isolate all the ultraviolet divergences (UV) in the one-loop
amplitudes generated by FeynArts \cite{feynart} and remove them with
the on-mass-shell renormalization scheme \cite{on-shell}. The
FormCalc-6.1 \cite{formcalc} and LoopTools-2.5 \cite{looptools} are
employed to simplify the amplitudes and to perform the numerical
calculations respectively.

On the other hand, the infrared (IR) divergences arising from the
contributions of virtual gluon exchange in loops are still left.
According to the Kinoshita-Lee-Nauenberg (KLN) theorem \cite{kln},
these IR divergences will be canceled by the real gluon
bremsstrahlung corrections in the soft gluon limit. We denote the
momentums of initial and final states for the real gluon emission
process as follows:
\begin{equation}
e^+(p_1)+e^-(p_2) \to t(k_1)+\bar{t}(k_2)+\gamma(k_3)+g(k) \, .
\end{equation}
We take the phase-space-slicing method to isolate the IR singularity
and divide the real corrections into the hard and soft parts by
the energy of the emitted gluon in the
calculations \cite{phase-slice1,phase-slice2}:
\begin{equation}
\Delta \sigma_{real}=\Delta \sigma_{ soft}+\Delta \sigma_{hard} \, .
\end{equation}
In the soft gluon approximation \cite{soft-photon}, we can obtain the
soft part of the cross sections by the following equation:
\begin{eqnarray}
\label{s} {d} \Delta\sigma_{soft} = d \sigma_{{0}}
\frac{\alpha_{s}C_F}{2 \pi^2} \int_{E_{g} \leq \Delta E_{g}}
\frac{d^3 \vec{k}}{2 E_{g}} \left( \frac{k_1}{k_1 \cdot k} -
\frac{k_2}{k_2 \cdot k} \right)^2,
\end{eqnarray}
where $C_{F}=\frac{4}{3}$, $E_g = \sqrt{|\vec{k}|^2+m_g^2}$ and we
give a fictitious mass $m_g$ to the gluon to eliminate the IR
divergence. Note that this dependence on the non-physical mass will
be exactly canceled by the virtual corrections which are also
evaluated with a non-zero gluon mass. $\Delta E_g$ is the energy
cutoff of the soft gluon and we require $E_g \leq \Delta E_g \ll
\sqrt{s}/2$. The hard gluon ($E_g \geq \Delta E_g$) radiation
corrections, which are insensitive to the small gluon mass, can be
directly evaluated by the numerical Monte Carlo method \cite{vegas}.

Finally, the finite total cross section of the process $e^{+}e^{-}
\to t \bar{t}\gamma$ including the LO ($\sigma_0$) and NLO QCD
corrections ($\Delta\sigma_{QCD}$) can be expressed as
\begin{equation}\label{cs}
\sigma_{tot}  = \sigma_0 + \Delta \sigma_{QCD} = \sigma_0 +
\Delta\sigma_{vir} + \Delta\sigma_{soft} + \Delta\sigma_{hard} =
\sigma_0(1 + \delta_{QCD})\, ,
\end{equation}
where $\delta_{QCD}\equiv\Delta\sigma_{QCD}/\sigma_{0}$ is the
relative QCD corrections at the order of ${\cal O}(\alpha_s)$.

\section{numerical results and discussions}

%%%%%%%%%%%%%% Fig.1 %%%%%%%%%%%%%%%%%%%%%%%%%%%%%%%%%%%%%%%%
\begin{figure}[htbp]
\includegraphics[width=3.2in,height=2.8in]{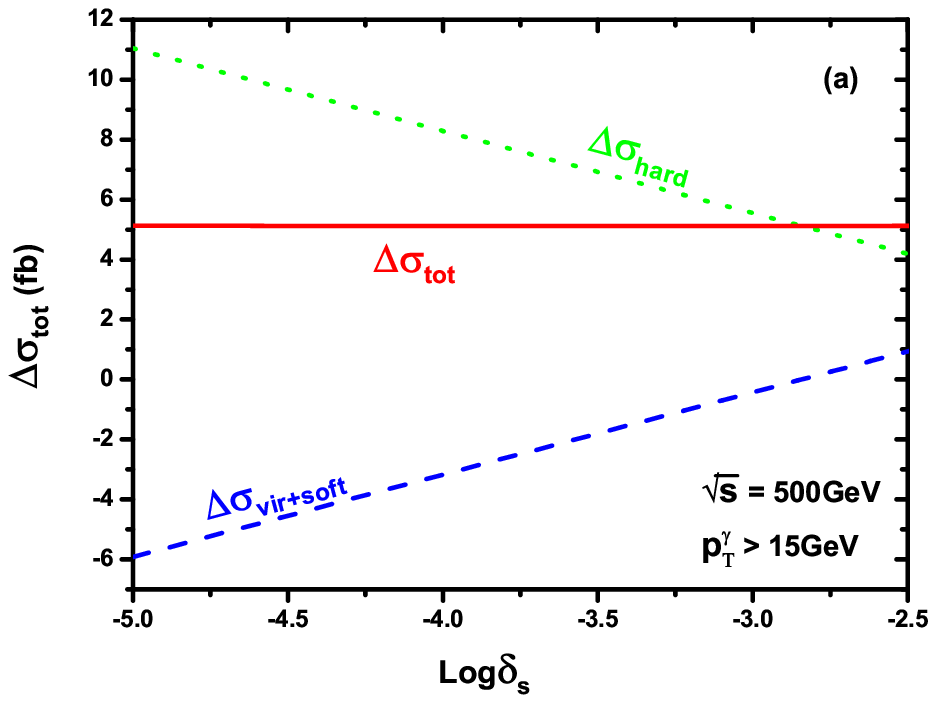}
\vspace{0in}%
\includegraphics[width=3.2in,height=2.8in]{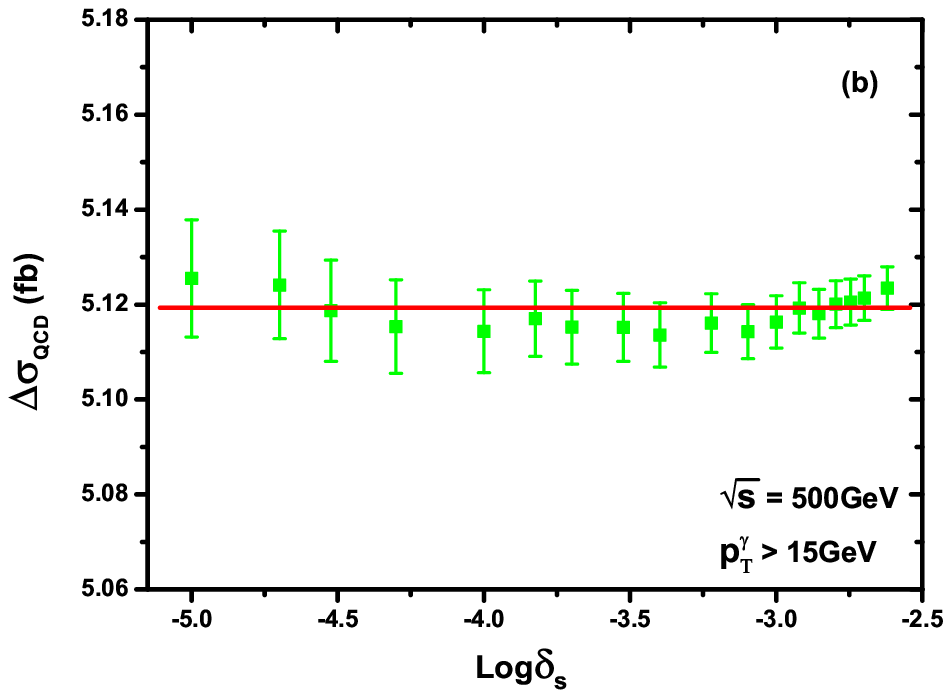}
\hspace{-0.1in}%
\vspace{-0.5cm}\caption{(a) The dependence of the NLO QCD
corrections on the soft cutoff $\delta_s$ in the process $e^{+}e^{-}
\to t\bar{t}\gamma$ for $m_g=10^{-8}$ GeV at $\sqrt{s}=500$GeV; (b)
The amplified curve marked with the calculation errors for
$\Delta\sigma_{QCD}$ versus $\delta_s$.}
\end{figure}

In the numerical evaluations, we take the input parameters of the SM
as \cite{pdg}
\begin{eqnarray}
m_t=171.2{\rm ~GeV}, ~~m_{e}=0.519991{\rm ~MeV}, ~~m_{Z}=91.19 {\rm
~GeV},\nonumber
\\~~\sin^{2}\theta_W=0.2228, ~~\alpha(m_Z^2)^{-1}=127.918~~~~~~~~~~~~~~~~
\end{eqnarray}
Since the total cross section is independent of the non-physical
parameters $m_g$ and the soft cutoff $\delta_s (\delta_s\equiv
\Delta E_{g}/E_{b},E_{b}=\sqrt{s}/2)$, we display the curves of the
NLO QCD corrections versus the cutoff $\delta_s$ for
$m_g=10^{-8}{\rm ~GeV}$ in Fig.1(a), where we fix the
renormalization scale $\mu=\mu_{0}=m_t$. For the strong coupling
constant $\alpha_s(\mu)$, we use the two-loop evolution of it with
the QCD parameter $\Lambda^{n_{f}=5}=226{\rm ~MeV}$ and get
$\alpha_s(\mu_0)=0.1078$. It can be seen that the values of
$\Delta\sigma_{hard}$ and $\Delta\sigma_{vir}+\Delta\sigma_{soft}$
vary with the change of the soft cutoff $\delta_s$, but the total
NLO QCD correction $\Delta\sigma_{QCD}$ is not dependent on
$\delta_s$ within the reasonable calculation errors. In order to
demonstrate this more clearly, we amplify the curve of
$\Delta\sigma_{QCD}$ in Fig.1(b). We also verify that the total
correction is indeed independent on $m_g$ for the fixed $\delta_s$.
Therefore, in the following calculations, we take
$\delta_s=2\times10^{-3}$ and $m_g=10^{-8}{\rm ~GeV}$.

%%%%%%%%%%%%%% Fig.2 %%%%%%%%%%%%%%%%%%%%%%%%%%%%%%%%%%%%%%%%
\begin{figure}[htbp]
\includegraphics[width=3.2in,height=2.8in]{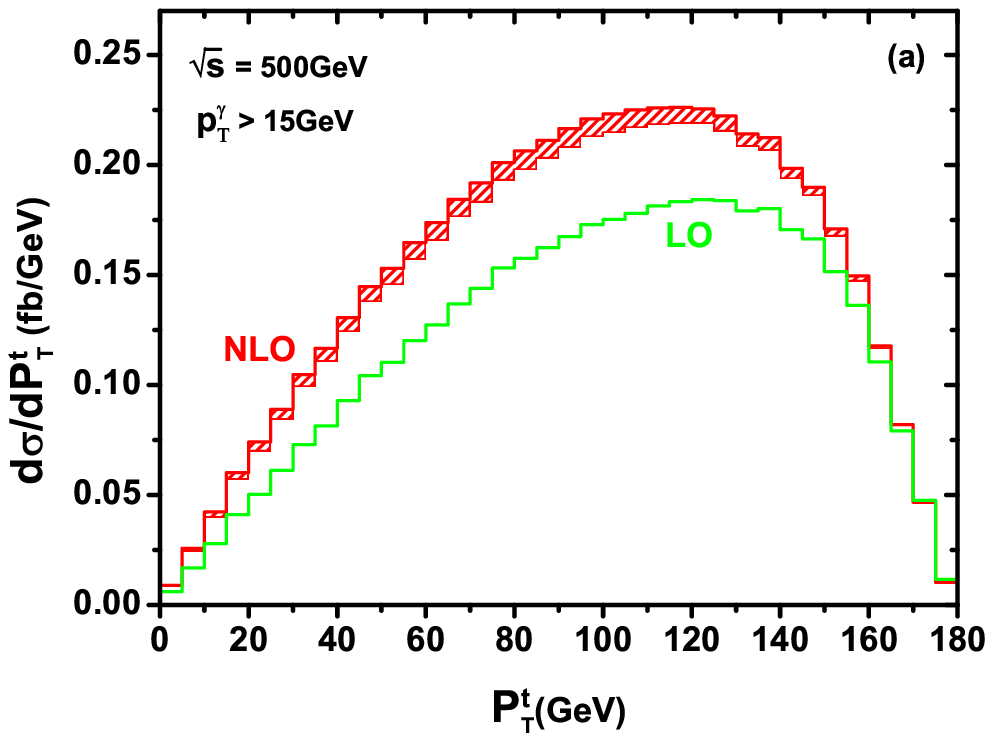}
\includegraphics[width=3.2in,height=2.8in]{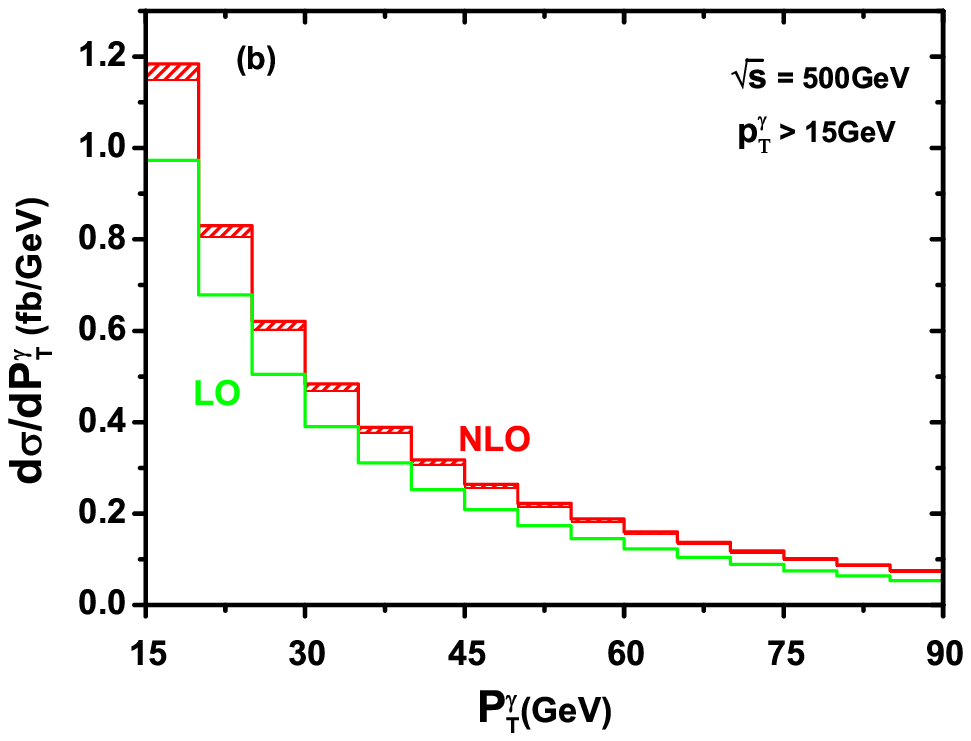}
\hspace{-0.1in}%
\vspace{-0.5cm}\caption{The transverse momentum distributions of the
top quark and photon at LO and NLO QCD respectively for the process $e^{+}e^{-}
\to t\bar{t}\gamma$. The bands correspond to the variation of the
renormalization scale in the interval $\mu_{0}/2 < \mu < 2\mu_{0}$.}
\end{figure}

In Fig.2(a-b), we show the transverse momentum distributions of the top
quark and photon at LO and NLO QCD for $\sqrt{s}=500{\rm ~GeV}$. In
the calculations, we maintain the electron mass and impose a
transverse momentum cut $p^{\gamma}_{T}>15{\rm ~GeV}$ to exclude
soft photon emission. It can be seen that the QCD corrections
greatly enhance the magnitudes of the LO differential cross section
$d\sigma_{0}/dp_{T}$. But the shapes of these distributions are not
dramatically changed. Most of the top quarks are produced in the region
of $40{\rm ~GeV}<p^{t}_{T}<150{\rm ~GeV}$; while the photons are
inclined to distribute in the region $15{\rm
~GeV}<p^{\gamma}_{T}<60{\rm ~GeV}$. We note that the detection of
energetic photons produced by hard scattering goes through the
definition of an isolation criterion \cite{isolation}. However, this
relies on the detailed Monte Carlo simulation, such as parton
shower, which is beyond the scope of our study. In table I, we
present the effects of different $p^{\gamma}_{T}$ cuts on the LO
cross section, the NLO QCD total cross sections and the $K$-factor of the
$t\bar{t}\gamma$ production at $\sqrt{s}=500$ GeV. For a higher
$p^{\gamma}_T$ cut, the cross sections become smaller and the
relative corrections get larger.

\begin{table}
\begin{tabular}{|c|c|c|c|}
\hline $p_T^\gamma$ (GeV)  & $\sigma_{0}(fb)$
& $\sigma_{tot}(fb)$ & $K$ \\
\hline 10  & 29.37(3)   & 35.96(3)   & 1.224(1)  \\
\hline 15  & 21.62(2)   & 26.76(2)   & 1.238(1)  \\
\hline 20  & 16.76(2)   & 20.94(2)   & 1.249(2)  \\
\hline 25  & 13.37(1)   & 16.87(1)   & 1.262(1)  \\
\hline 30  & 10.84(1)   & 13.81(1)   & 1.274(1)  \\
\hline
\end{tabular}
\caption{The LO cross sections, the NLO QCD total cross sections
and $K$-factors under different $p^{\gamma}_{T}$ cuts for process
$e^{+}e^{-} \to t\bar{t}\gamma$ at $\sqrt{s}=500$GeV.}
\end{table}

%%%%%%%%%%%%%% Fig.3 %%%%%%%%%%%%%%%%%%%%%%%%%%%%%%%%%%%%%%%%
\begin{figure}[htbp]
\includegraphics[width=3.2in,height=2.8in]{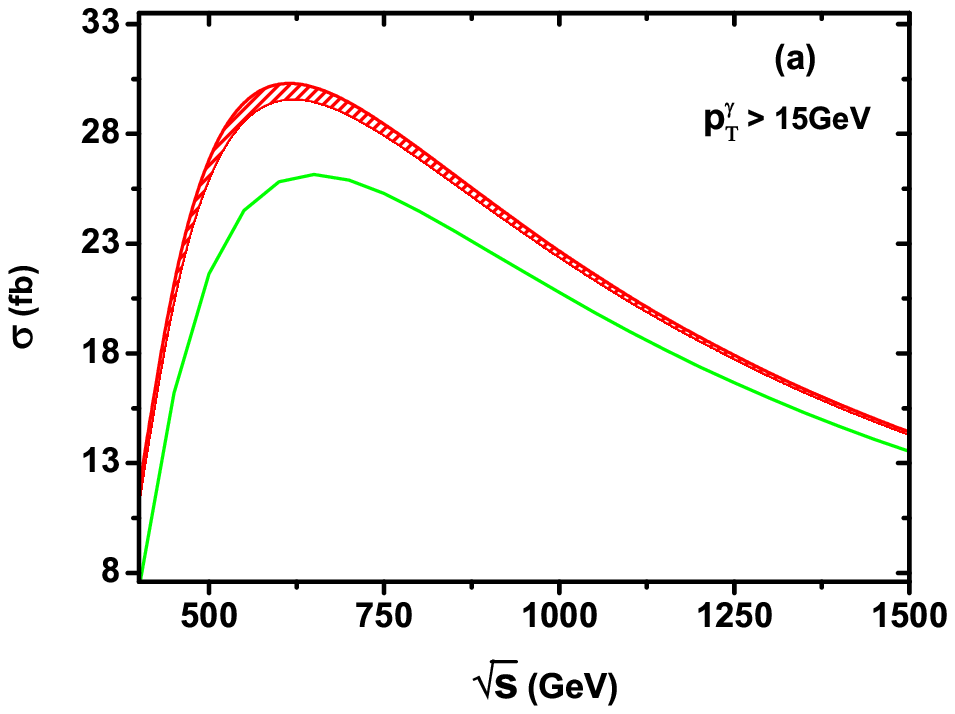}
\vspace{0in}%
\includegraphics[width=3.2in,height=2.8in]{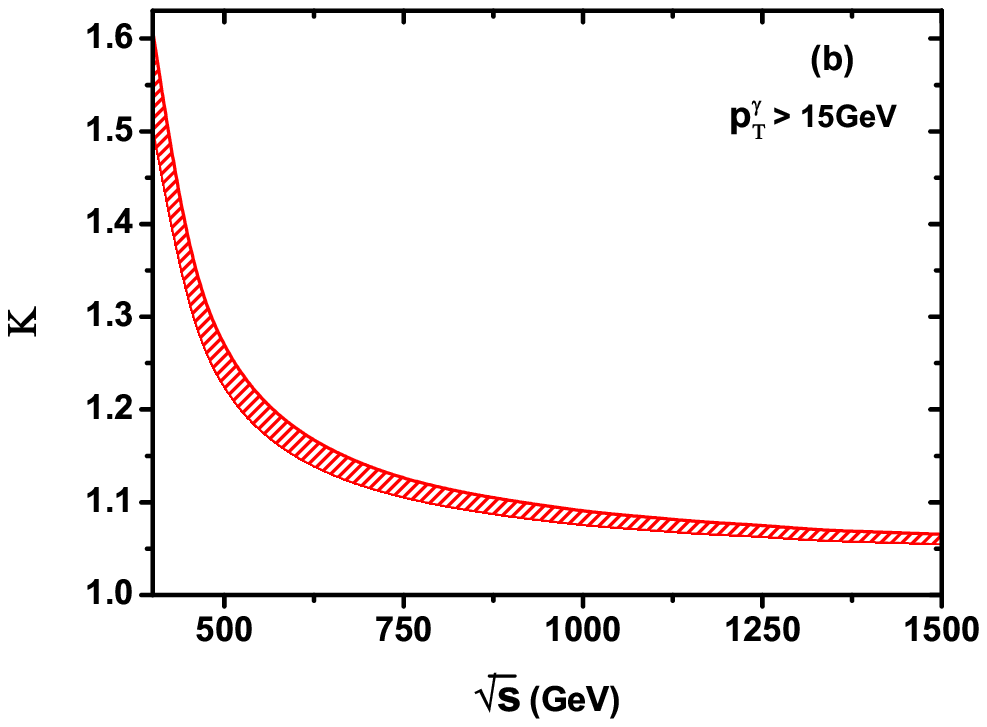}
\hspace{-0.1in}%
\vspace{-0.5cm}\caption{ (a) The cross sections of $e^{+}e^{-} \to
t\bar{t}\gamma$ versus $\sqrt{s}$ at LO and NLO QCD
respectively; (b) The corresponding $K$-factor versus $\sqrt{s}$.
The bands correspond to the variation of the renormalization scale in
the interval $\mu_{0}/2 < \mu < 2\mu_{0}$.}
\end{figure}

\begin{table}
\bc
\begin{tabular}{l r@{.}l r@{.}l r@{.}l r@{.}l r@{.}l r@{.}l r@{.}l r@{.}l r@{.}l}
\hline
 & \multicolumn{6}{c|}{$\sqrt{s}=500\gev$} & \multicolumn{6}{c|}{$\sqrt{s}=800\gev$} & \multicolumn{6}{c}{$\sqrt{s}=1500\gev$}\\
 \hline
$\mu$ &\multicolumn{2}{c}{ $\sigma_{tot} (fb)$}
&\multicolumn{2}{c}{$K$} &\multicolumn{2}{c}{$A_{FB}^{t(tot)} (\%)$}
&\multicolumn{2}{c}{ $\sigma_{tot} (fb)$} &\multicolumn{2}{c}{$K$}
&\multicolumn{2}{c}{$A_{FB}^{t(tot)} (\%)$} &\multicolumn{2}{c}{
$\sigma_{tot} (fb)$} &\multicolumn{2}{c}{$K$}
&\multicolumn{2}{c}{$A_{FB}^{t(tot)} (\%)$}\\
\hline
$\mu_0/2$  & ~27&30(2)  & ~1&263(1)  & 45&72(4)  & 27&33(3)  & ~1&116(1)  & 55&3(2)  & 14&41(3)  & ~~1&065(2)  & 55&9(3)\\
$\mu_0$   & ~26&76(2)  & ~1&238(1)  & 45&82(4)  & 27&06(3)  &  ~1&105(1)  & 55&4(2)  & 14&34(3)  & ~~1&060(2)  & 56&0(3)\\
$2\mu_0$  & ~26&31(2)  & ~1&217(1)  & 45&90(5)  & 26&83(3)  &  ~1&096(1)  & 55&5(6)  & 14&26(3)  & ~~1&054(2)  & 56&1(3)\\
\hline
\end{tabular}\ec
\caption{\label{tab2} The numerical results of the NLO QCD total
cross sections, the $K$-factor and the top quark forward-backward
asymmetry at different values of the renormalizaiotn scale for the
process $e^{+}e^{-} \to t\bar{t}\gamma$.}
\end{table}

In Fig.3(a-b), we give the dependence of cross sections and relative
corrections on the center-of-mass energy $\sqrt{s}$. Since the
process $e^{+}e^{-} \to t\bar{t}\gamma$ in the SM is induced by the
pure electro-weak interaction at the order ${\cal O}(\alpha^3)$, the
LO cross section will not be affected by the variation of the
renormalization scale in the strong coupling. However, the NLO QCD
corrections are leading order in $\alpha_s$ and show a weak
dependence on the scale, due to the suppression of the loop factor.
We display the values of the NLO QCD total cross section
($\sigma_{tot}$), the $K$-factor and the top quark forward-backward
asymmetry at the renormalization scale $\mu=\mu_0/2,\mu_0,2\mu_0$ in
table II. The uncertainty of the NLO scale dependence is
approximately $3.7\%$ ($1.8\%$,$1.0\%$) for $\sqrt{s}=500$ GeV (800
GeV, 1500 GeV) when the scale $\mu$ is varied between $\mu_0/2$ and
$2\mu_0$. The uncertainty is defined as $\delta=[|\sigma(\mu_0/2) -
\sigma(\mu_0)| + |\sigma(2\mu_0) - \sigma(\mu_0)|]/\sigma(\mu_0)$.
When setting the scale at $\mu_0$, we find that the largest
production rates of $t\bar{t}\gamma$ will reach about $25.82$ fb and
$29.98$ fb at LO and NLO QCD respectively around $\sqrt{s}=600 {\rm
~GeV}$, where the threshold effect may dominate. The corresponding
relative QCD correction can be $16.1\%$. When $\sqrt{s}$ is greater
than 600 GeV, the cross sections drop rapidly, due to the
$s$-channel suppression.

%%%%%%%%%%%%%% Fig.4 %%%%%%%%%%%%%%%%%%%%%%%%%%%%%%%%%%%%%%%%
\begin{figure}[htbp]
\includegraphics[width=3.2in,height=2.8in]{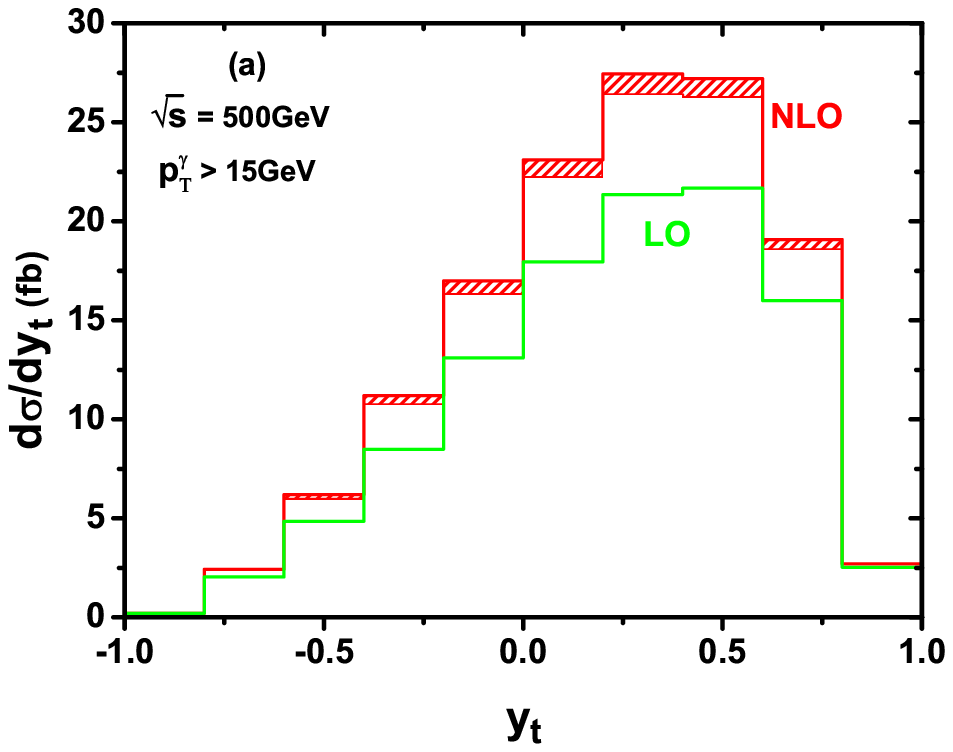}
\includegraphics[width=3.2in,height=2.8in]{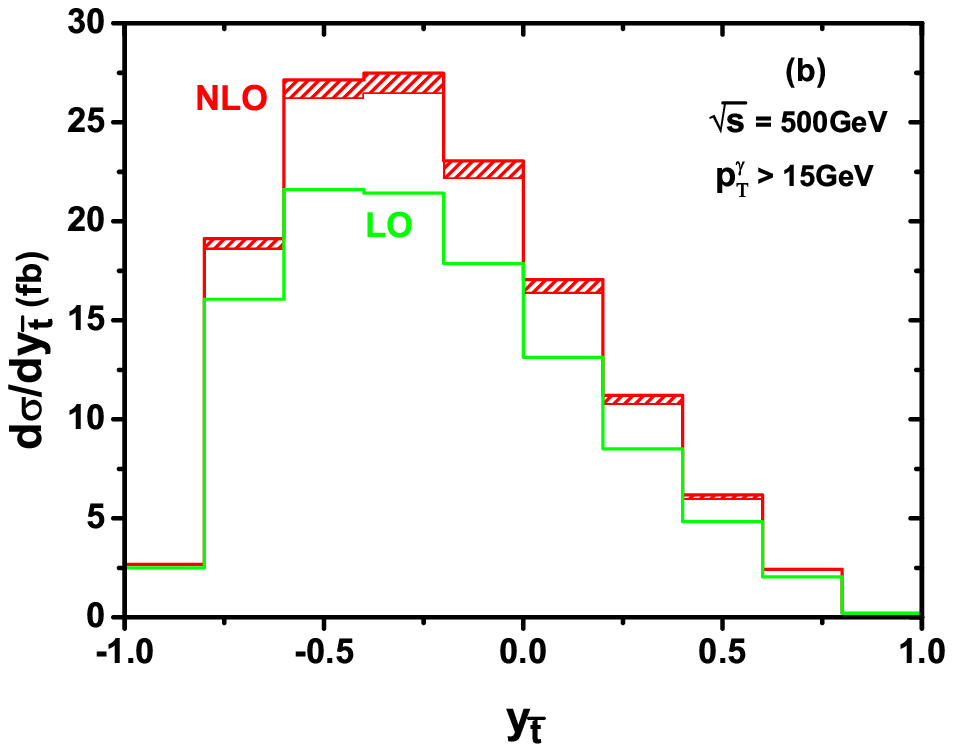}
\hspace{-0.1in}%
\vspace{-0.5cm}\caption{The rapidity distributions of top and
anti-top at LO and NLO QCD respectively in the process $e^{+}e^{-}
\to t\bar{t}\gamma$. The bands correspond to the variation of the
renormalization scale in the interval $\mu_{0}/2 < \mu < 2\mu_{0}$.}
\end{figure}

We also investigate the rapidity-differential cross sections of the
top (anti-top) quark in Fig.4(a-b) and find that the events of the
top (anti-top) quark for $y>0$ are more (less) than that for $y<0$.
This asymmetry is caused by the huge interference effect between the
photon and $Z^{0}$ boson mediated in the process $e^{+}e^{-}\to
\gamma^{*}/Z^{0}\to t\bar{t}\gamma$ \cite{asymmetry}. The QCD
corrections enhance the LO distributions of the top (anti-top) quark
but do not distort their shapes significantly. In order to present
this asymmetry, we can define the top quark forward-backward
asymmetry ($A_{FB}^{t}$) in the process $e^{+}e^{-} \to t
\bar{t}\gamma$ as
\begin{eqnarray}
A_{FB}^t=\frac{N(y_t>0)-N(y_t<0)}{N(y_t>0)+N(y_t<0)} .
\end{eqnarray}
Here $N(y_t>0)$ and $N(y_t<0)$ denote the events of top quarks
moving along or against a given direction, which is chosen as the
direction of the incoming particle $e^{-}$ in our calculations. In
Fig.5, we can see that the QCD corrections give arise to a negative
contribution to the LO forward-backward asymmetry. We list the
values of $A^{t}_{FB}$ at different scales in table.II. We find that
the dependence of $A^{t}_{FB}$ on the renormalization scale is very
weak, due to the cancelation of strong coupling between numerator
and denominator in Eq.(6). It is also noted that the value of this
asymmetry is not sensitive to the collision energy when $\sqrt{s}$
is greater than 900 {\rm ~GeV}. The maximal values of $A^{t}_{FB}$
can reach $58.1\%$ and $56.4\%$ for $\mu=\mu_0$ at LO and NLO QCD
respectively.

%%%%%%%%%%%%%%%%%%%%    fig.5   %%%%%%%%%%%%%%%%%%%%%%%%%%%%
\begin{figure}[htb]
\epsfig{file=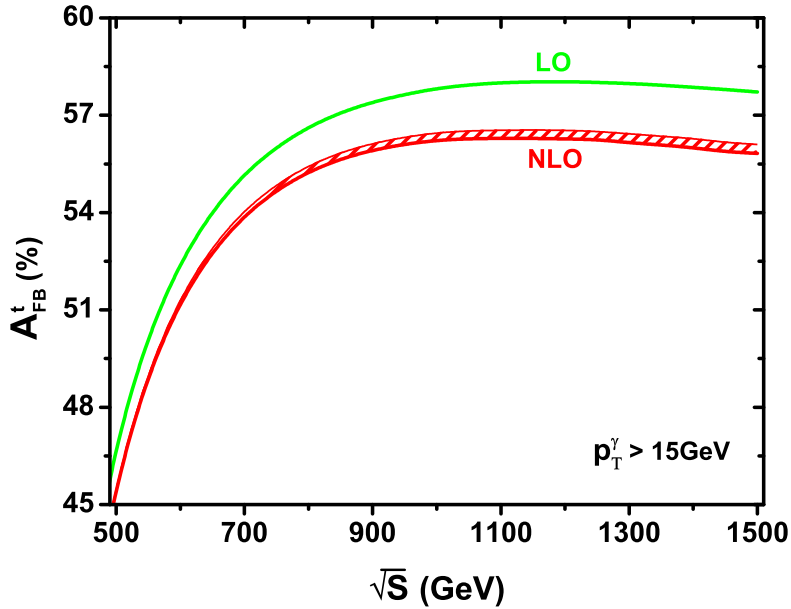,width=12cm} \vspace{-0.5cm} \caption{The
dependence of the forward-backward asymmetry $A^{t}_{FB}$ on $\sqrt{s}$
in the production of $t\bar{t}\gamma$ at LO and NLO QCD respectively
at the ILC. The band corresponds to the variation of the
renormalization scale in the interval $\mu_{0}/2 < \mu < 2\mu_{0}$.}
\label{fig1}
\end{figure}
%%%%%%%%%%%%%%

\section{Conclusions}
In this paper, we discussed in detail the one-loop QCD
corrections to the process $e^{+}e^{-} \to t\bar{t}\gamma$ at the
ILC. We found that the QCD corrections can significantly enhance the
production rate of $t\bar{t}\gamma$ and show a weak dependence
on the renormalization scale. The shapes of differential
distributions of the top quark and photon are not be greatly affected
by the QCD corrections. When fixing $\mu=\mu_0$, we found that the
total cross section and the top quark forward-backward asymmetry can
respectively reach $26.76$ ($27.06$,~$14.34$) $fb$ and
$45.82$ ($55.4$,~$56.0$)\% at NLO QCD for $\sqrt{s}=500~(800,~1500)$ GeV.

\section*{Acknowledgement}
The work is supported by the Startup Foundation for Doctors of Henan
Normal University under contract No.11112.


\begin{thebibliography}{99}

\bibitem{SM}
S. L. Glashow, Nucl. Phys. {\bf 22}, 579 (1961); S. Weinberg,
Phys.Rev. Lett. {\bf 19}, 1264 (1967); H. D. Politzer, Phys. Rep.
{\bf 14}, 129 (1974).

\bibitem{higgs}
P. W. Higgs, Phys. Lett {\bf 12}, 132 (1964) Phys. Rev. Lett. {\bf
13}, 508 (1964); Phys. Rev. {\bf 145}, 1156 (1966); F. Englert and
R.Brout, Phys. Rev. Lett. {\bf 13}, 321 (1964); G. S. Guralnik, C.
R. Hagen and T. W. B. Kibble, Phys. Rev. Lett. {\bf 13} 585, (1964);
T. W. B. Kibble, Phys. Rev. {\bf 155} 1554, (1967).

\bibitem{top-exp1}
F. Abe, {\it et al.} (CDF Collaboration), Phys.  Rev. Lett. {\bf
74}, 2626 (1995); S. Abachi, {\it et al.} (D\O\ Collaboration),
Phys. Rev. Lett. {\bf 74}, 2632 (1995).

\bibitem{top-exp2}
M.~Lancaster, {\it et al.}
%``Combination of CDF and D0 results on the mass of the top quark using up to
%5.8~fb-1 of data,''
arXiv:1107.5255 [hep-ex];
%%CITATION = ARXIV:1107.5255;%%
V.~M.~Abazov {\it et al.} [ D0 Collaboration ],
%``Determination of the width of the top quark,''
Phys.\ Rev.\ Lett.\  {\bf 106}, 022001 (2011).

\bibitem{top-decay}
I.~I.~Y.~Bigi, Y.~L.~Dokshitzer, V.~A.~Khoze, J.~H.~Kuhn and P.~M.~Zerwas,
%``Production and Decay Properties of Ultraheavy Quarks,''
Phys.\ Lett.\  B {\bf 181}, 157 (1986).
%%CITATION = PHLTA,B181,157;%%

\bibitem{top-rev} For top quark reviews, see, e.g.,
               W. Bernreuther, \JPG 35, 083001,(2008)
               D. Chakraborty, J. Konigsberg, D. Rainwater,
               {\it Ann. Rev. Nucl. Part. Sci.} {\bf 53}, 301  (2003);
               E.~H.~Simmons, hep-ph/0211335;
                 %%CITATION = HEP-PH 0211335;%%
               C.-P. Yuan,  hep-ph/0203088;
                 %%CITATION = HEP-PH 0203088;%%
               S. Willenbrock, hep-ph/0211067;
                 %%CITATION = HEP-PH 0211067;%%
               M. Beneke, {\it et al.}, hep-ph/0003033;
                 %%CITATION = HEP-PH 0003033;%%
               T. Han, arXiv:0804.3178;
                 %%CITATION = ARXIV:0804.3178;%%
               For model-independent new physics study, see, e.g.,
               C. T. Hill and S. J. Parke, \PRD49, 4454 (1994);
                  %%CITATION = PHRVA,D49,4454;%%
               K. Whisnant, {\it et al.},  \PRD56, 467 (1997);
                  %%CITATION = PHRVA,D56,467;%%
               J. M. Yang, B.-L. Young, \PRD56, 5907 (1997);
                  %%CITATION = PHRVA,D56,5907;%%
               K. Hikasa, {\it et al.}, \PRD58, 114003 (1998);
                  %%CITATION = PHRVA,D58,114003;%%
               J. A. Aguilar-Saavedra, arXiv:0811.3842;
                 %%CITATION = ARXIV:0811.3842;%%
               R.A. Coimbra, {\it et al.}, arXiv:0811.1743.
                 %%CITATION = ARXIV:0811.1743;%%

\bibitem{top-afb}
T.~Aaltonen, {\it et al.}  [The CDF Collaboration], \PRD83, 112003
(2011); Y.~Takeuchi, {\it et al.},
http://www-cdf.fnal.gov/physics/new/top/2011/DilAfb/Note 10398;
V.~M. Abazov, et~al. [The D0 Collaboration], arXiv:1107.4995.

\bibitem{top-exp3}
  G.~Aad {\it et al.}  [ATLAS Collaboration],
  %``Measurement of the top quark pair production cross section in pp collisions
  %at sqrt(s) = 7 TeV in dilepton final states with ATLAS,''
  arXiv:1108.3699 [hep-ex];
  %%CITATION = ARXIV:1108.3699;%%
  S.~Chatrchyan {\it et al.}  [CMS Collaboration],
  %``Measurement of the Top-antitop Production Cross Section in pp Collisions at
  %sqrt(s)=7 TeV using the Kinematic Properties of Events with Leptons and
  %Jets,''
  arXiv:1106.0902 [hep-ex];
  %%CITATION = ARXIV:1106.0902;%%
  S.~Chatrchyan {\it et al.}  [CMS Collaboration],
  %``Measurement of the t-channel single top quark production cross section in
  %pp collisions at sqrt(s) = 7 TeV,''
  arXiv:1106.3052 [hep-ex];
  %%CITATION = ARXIV:1106.3052;%%
F.~Deliot and D.~Glenzinski,
%``Top Quark Physics at the Tevatron,''
arXiv:1010.1202 [hep-ex];
%%CITATION = ARXIV:1010.1202;%%
M.~A.~Pleier,
%``Review of Top Quark Properties Measurements at the Tevatron,''
Int.\ J.\ Mod.\ Phys.\  A {\bf 24}, 2899 (2009); B.~Stelzer,
%``Review of Top Quark Measurements,''
arXiv:1004.5368 [hep-ex].


\bibitem{CDF-ttr}
T.~Aaltonen {\it et al.} (CDF Collaboration), [arXiv:1106.3970
[hep-ex]].

\bibitem{top-exotic-charge}
U.~Baur, M.~Buice and L.~H.~Orr,
%``Direct measurement of the top quark charge at hadron colliders,''
Phys.\ Rev.\  D {\bf 64}, 094019 (2001).
%%CITATION = PHRVA,D64,094019;%%


\bibitem{LHC-ttr}  M. Ciljak {\it et al.}, ATLAS Note PHYS-2003-35 (2003)

\bibitem{ttr-hadron}
D.~Peng-Fei, M.~Wen-Gan, Z.~Ren-You, H.~Liang, G.~Lei and
W.~Shao-Ming, Phys.\ Rev.\ D80, 014022 (2009); K.~Melnikov,
M.~Schulze and A.~Scharf, Phys.\ Rev.\  D {\bf 83}, 074013 (2011)

\bibitem{top-ilc}  http://www.linearcollider.org/about/Publications/Reference-Design-Report, volume II.

\bibitem{top-gauge-indep}
J.~L.~Hewett,
%``Probing top quark couplings at lepton and photon colliders,''
Int.\ J.\ Mod.\ Phys.\  {\bf A13}, 2389-2398 (1998); B.~Grzadkowski,
Z.~Hioki {\it et al.},
%``Probing anomalous top quark couplings induced by dimension-six operators at
%photon colliders,''
Nucl.\ Phys.\  B {\bf 689}, 108 (2004);
%``New-physics search through gamma gamma ---> t anti-t ---> lX X,''
Nucl.\ Phys.\ Proc.\ Suppl.\  {\bf 157}, 246 (2006);
%%CITATION = NUPHZ,157,246;%%
J.~A.~Aguilar-Saavedra,
%``A Minimal set of top anomalous couplings,''
Nucl.\ Phys.\  B {\bf 812}, 181 (2009).
%%CITATION = NUPHA,B812,181;%%

\bibitem{top-lh}
  N.~Arkani-Hamed {\it et al.},
  %``The Littlest Higgs,''
  JHEP {\bf 0207}, 034 (2002);
  %%CITATION = JHEPA,0207,034;%%
  T.~Han {\it et al.},
  %``Phenomenology of the little Higgs model,''
  Phys.\ Rev.\  D {\bf 67}, 095004 (2003);
  %%CITATION = PHRVA,D67,095004;%%


\bibitem{top-cp}
P.~Poulose and S.~D.~Rindani,Phys.\ Rev.\  D {\bf 57}, 5444 (1998);
[Erratum-ibid.\  D {\bf 61}, 119902 (2000)]; Phys.\ Rev.\  D {\bf
57}, 5444 (1998) [Erratum-ibid.\  D {\bf 61}, 119902 (2000)];
H.~Y.~Zhou, Phys.\ Lett.\  B {\bf 439}, 393 (1998); W.~Hollik {\it
et al.} , Nucl.\ Phys.\ B {\bf 551}, 3 (1999) [Erratum-ibid.\ B {\bf
557}, 407 (1999)]; W.~Hollik {\it et al.}, Phys.\ Lett.\  B {\bf
425}, 322 (1998); D.~Atwood {\it et al.}, Phys.\ Rept.\ {\bf 347}, 1
(2001);  N.~Liu {\it et al.}, Phys.\ Rev.\  D {\bf 82}, 015009
(2010); T.~Ibrahim and P.~Nath, Phys.\ Rev.\  D {\bf 82}, 055001
(2010).

\bibitem{tt-ilc}
  J.~Jersak {\it et al.},
  %``Electroweak Production of Heavy Quarks in e+ e- Annihilation,''
  Phys.\ Rev.\  D {\bf 25}, 1218 (1982)
  [Erratum-ibid.\  D {\bf 36}, 310 (1987)]
  [Phys.\ Rev.\  D {\bf 36}, 310 (1987)];
  %%CITATION = PHRVA,D36,310;%%
  B.~Kamal {\it et al.},
   %``Heavy quark production by polarized and unpolarized photons in
  %next-to-leading order,''
  Phys.\ Rev.\  D {\bf 51}, 4808 (1995)
  [Erratum-ibid.\  D {\bf 55}, 3229 (1997)]
  [Phys.\ Rev.\  D {\bf 55}, 3229 (1997)];
  %%CITATION = PHRVA,D55,3229;%%
  W.~Beenakker {\it et al.},
  %``e+ e- annihilation into heavy fermion pairs at high-energy colliders,''
  Nucl.\ Phys.\  B {\bf 365}, 24 (1991);
  %%CITATION = NUPHA,B365,24;%%
  A.~A.~Akhundov {\it et al.},
   %``QED radiative corrections to massive fermion production in e+ e-
  %annihilation,''
  Phys.\ Lett.\  B {\bf 261}, 321 (1991);
  %%CITATION = PHLTA,B261,321;%%
  A.~Denner {\it et al.},
   %``Radiative corrections to $\gamma \gamma \to t \bar{t}$ in the electroweak
  %standard model,''
  Phys.\ Rev.\  D {\bf 53}, 44 (1996).
  %%CITATION = PHRVA,D53,44;%%

\bibitem{ttcp-ee}
D.~Chang {\it et al.}, Nucl.\ Phys.\  B {\bf 408}, 286 (1993)
[Erratum-ibid.\  B {\bf 429}, 255 (1994)]; W. Bernreuther {\it et
al.}, Phys.\ Lett.\ B 279, 389 (1992); M.S. Baek {\it et al.},
Phys.\ Rev.\ D56, 6835 (1997).

\bibitem{ttcp-rr}
S.Y. Choi, and Hagiwara, Phys.\ Lett.\ B359, 369(1995); L.~Han {\it
et al.}, Phys.\ Rev.\  {\bf D56}, 265-275 (1997); W.~-G.~Ma {\it et
al.}, Commun.\ Theor.\ Phys.\ {\bf 26}, 455-460 (1996); Commun.\
Theor.\ Phys.\ {\bf 27}, 101-104 (1997); M.~-L.~Zhou {\it et al.},
J.\ Phys.\ G {\bf G25}, 27-43 (1999).

\bibitem{bbr-lep}
L.~Magnea and E.~Maina, Phys.\ Lett.\  B {\bf 385}, 395 (1996).

\bibitem{feynart}
T. Hahn, Comput. Phys. Commun. 140, 418 (2001).

\bibitem{on-shell}
  M.~Bohm, H.~Spiesberger and W.~Hollik,
   %``On the One Loop Renormalization of the Electroweak Standard Model and Its
  %Application to Leptonic Processes,''
  Fortsch.\ Phys.\ 34 (1986) 687;
  %%CITATION = FPYKA,34,687;%%
  W.~F.~L.~Hollik,
   %``Radiative Corrections in the Standard Model and their Role for Precision
  %Tests of the Electroweak Theory,''
  Fortsch.\ Phys.\ 38 (1990) 165;
  %%CITATION = FPYKA,38,165;%%}
  B.~Grzadkowski and W.~Hollik,
   %``Radiative corrections to the top quark width within two Higgs doublet
  %models,''
  Nucl.\ Phys.\ B 384 (1992) 101.
  %%CITATION = NUPHA,B384,101;%%

\bibitem{formcalc}
T. Hahn, M. Perez-Victoria, Comput. Phys. Commun. 118, 153 (1999).

\bibitem{looptools}
G. J. van Oldenborgh, Phys Commun 66 (1991) 1, NIKHEF-H-90-15; G.¡¯t
Hooft and M. Veltman, Nucl. Phys. B153, 365 (1979); A. Denner,
Fortschr. Phys. 41, 307 (1993).


\bibitem{kln}
T. Kinoshita, J. Math. Phys. 3(1962) 650; T.D. Lee and M. Nauenberg,
Phys. Rev. 133(1964) 1549.

\bibitem{phase-slice1}
B. W. Harris and J.F. Owens, Phys. Rev. D65, 094032(2002);

\bibitem{phase-slice2}
W. T. Giele and E. W. N. Glover, Phys. Rev. D46, 1980 (1992); W. T.
Giele, E. W. Glover and D. A. Kosower, Nucl. Phys. B403, 633 (1993);
S. Keller and E. Laenen, Phys. Rev. D59, 114004 (1999).

\bibitem{soft-photon}
S. Dawson and L. Reina, Phys. Rev. D59, 054012 (1999).

\bibitem{vegas}
G.P. Legage, J. Comput. Phys. 27, 192(1978).

\bibitem{pdg}
C. Amsler {\it et al.}, Particle Data Group, \PLB667, 1 (2008).

\bibitem{isolation}

  E.~L.~Berger and J.~w.~Qiu,
  Phys.\ Lett.\  B {\bf 248}, 371 (1990);
  %%CITATION = PHLTA,B248,371;%%
  E.~W.~N.~Glover and W.~J.~Stirling,
  Phys.\ Lett.\  B {\bf 295}, 128 (1992);
  %%CITATION = PHLTA,B295,128;%%
  S.~Catani, M.~Fontannaz and E.~Pilon,
  Phys.\ Rev.\  D {\bf 58}, 094025 (1998);
  %%CITATION = PHRVA,D58,094025;%%
  S.~Frixione and W.~Vogelsang,
  Nucl.\ Phys.\  B {\bf 568}, 60 (2000).
  %%CITATION = NUPHA,B568,60;%%

\bibitem{asymmetry}
  M. Bohm {\it et al.}, CERN-TH-5536-89;
  G.~L.~Kane, G.~A.~Ladinsky and C.~P.~Yuan,
  % ``Using the Top Quark for Testing Standard Model Polarization and CP
  %Predictions,''
  Phys.\ Rev.\  D {\bf 45}, 124 (1992).
  %%CITATION = PHRVA,D45,124;%%



\end{thebibliography}
\end{document}